\documentclass[12pt]{article}
\usepackage{amsfonts}
\usepackage{amsmath}
\usepackage{amssymb}
\usepackage{graphicx}
\usepackage{color}
\usepackage[all, knot]{xy}
\usepackage{tikz}

\usepackage{ulem}

\usepackage[utf8]{inputenc}
\usepackage{epstopdf}
\usepackage[footnotesize]{caption}
\usepackage{amsthm}
\usepackage{enumitem}
\usepackage{mathrsfs}

\usepackage[margin=3cm]{geometry}

\def \be {\begin{equation}}
\def \ee {\end{equation}}
\def \bea {\begin{eqnarray}}
\def \eea {\end{eqnarray}}
\def \nn {\nonumber}

\def \rr {\raise.35ex\hbox{\small $\prime$}\kern-.17em{\mbox{\large $\imath$}}}

\def \dels {\partial\kern-.6em /\kern.1em}
\def \As {{A\kern-.5em / \kern.5em}}
\def \Ds {D\kern-.7em / \kern.5em}

\def \ks {k\kern-.5em /}
\def \ls {l\kern-.5em /}

\newcommand{\ci}[1]{}

\newcommand{\ba}{\begin{eqnarray}}
\newcommand{\ea}{\end{eqnarray}}
\newcommand{\bal}{\begin{align}}
\newcommand{\eal}{\end{align}}
\newcommand{\bay}[1]{\left(\begin{array}{#1}}
\newcommand{\eay}{\end{array}\right)}

\newcommand{\hide}[1]{}

\newlist{axioms}{enumerate}{2}
\setlist[axioms,1]{label=\textbf{A\arabic{axiomsi}.}, ref=A\arabic{axiomsi}}
\setlist[axioms,2]{label=\textbf{A\arabic{axiomsi}\rlap{\myEnumCounter{axiomsii}}.},%
                   ref=A\arabic{axiomsi}\myEnumCounter{axiomsii},%
                   align=parleft,%
                   leftmargin=0em,%
                   itemsep=1.4ex,%
                   before={\stepcounter{axiomsi}}}
                   
  \usetikzlibrary{decorations.markings}

\begin{document}

\begin{titlepage}
\begin{center}

\textbf{\LARGE
Integrability and Spectral Form Factor in Chern-Simons Formulation
\vskip.3cm
}
\vskip .5in
{\large

Chen-Te Ma$^{a, b, c, d}$ \footnote{e-mail address: yefgst@gmail.com} and Hongfei Shu$^{e, f}$ \footnote{e-mail address: shuphy124@gmail.com 
} 
\\
\vskip 1mm
}
{\sl
$^a$
Guangdong Provincial Key Laboratory of Nuclear Science,\\
 Institute of Quantum Matter,
South China Normal University, Guangzhou 510006, Guangdong, China.
\\
$^b$
School of Physics and Telecommunication Engineering,\\ 
South China Normal University, Guangzhou 510006, Guangdong, China.
\\
$^c$
The Laboratory for Quantum Gravity and Strings,\\
Department of Mathematics and Applied Mathematics,
University of Cape Town, Private Bag, Rondebosch 7700, South Africa.
\\
$^d$
Department of Physics and Center for Theoretical Sciences,\\ 
National Taiwan University, Taipei 10617, Taiwan, R.O.C..
\\
$^e$
Nordita, KTH Royal Institute of Technology and Stockholm University,
 Roslagstullsbacken 23, SE-106 91 Stockholm, Sweden.
\\
$^f$
Department of Physics, Tokyo Institute of Technology, Tokyo, 152-8551, Japan.
}\\
\vskip 1mm
\vspace{40pt}
\end{center}
\newpage
\begin{abstract}
We study the integrability from the spectral form factor in the Chern-Simons formulation. 
The effective action in the higher spin sector was not derived so far.
Therefore, we begin from the SL(3) Chern-Simons higher spin theory. Then the dimensional reduction in this Chern-Simons theory gives the SL(3) reparametrization invariant Schwarzian theory, which is the boundary theory of an interacting theory between the spin-2 and spin-3 fields at the infrared or massless limit. We show that the Lorentzian SL(3) Schwarzian theory is dual to the integrable model, SL(3) open Toda chain theory. Finally, we demonstrate the application of open Toda chain theory from the SL(2) case. The numerical result shows that the spectral form factor loses the dip-ramp-plateau behavior, consistent with integrability. The spectrum is not a Gaussian random matrix spectrum. We also give an exact solution of the spectral form factor for the SL(3) theory. This solution provides a similar form to the SL(2) case for $\beta\neq 0$. Hence the SL(3) theory should also do not have a Gaussian random matrix spectrum.
\end{abstract}
\end{titlepage}

\section{Introduction}
\label{sec:1}
\noindent 
One direction of quantum gravity is the {\it first-order formulation} of the {\it 3-dimensional Einstein gravity theory}, which describes gravitation. This theory has been formulated by the 3-dimensional Chern-Simons theory with the SL(2) group \cite{Witten:1988hc}. Although the 3-dimensional Einstein gravity theory is not renormalizable, the {\it Chern-Simons formulation} avoids this problem,  which enables us to study quantum gravity theory from this perspective.
\\ 

\noindent 
The study in the exact boundary dual theory \cite{Coussaert:1995zp} of 3-dimensional Anti-de Sitter (AdS$_3$) Einstein gravity theory begins with the 2-dimensional Liouville theory. The Liouville theory is a 2-dimensional conformal field theory (CFT$_2$) and has a continuous spectrum without a normalizable vacuum. On the contrary, the AdS$_3$ Einstein gravity theory has a discrete spectrum with a normalizable vacuum. Therefore, the bulk theory contradicts the fact of the Liouville theory. Hence this exact study does not provide a reliable result.
\\

\noindent
More recently, the boundary theory of the AdS$_3$ Einstein gravity theory has been derived \cite{Cotler:2018zff}. The boundary theory appears as a doubled copy of the 2-dimensional Schwarzian theory, similar to the Schwarzian theory \cite{Cotler:2018zff}. Hence the holographic study is the nearly AdS/CFT correspondence. The problem of the exact study in the 2-dimensional Liouville theory is that the gauge fields have the SL(2) redundancy \cite{Cotler:2018zff}. Ones thus need to identify these field configurations to obtain additional constraints in the path integration \cite{Cotler:2018zff}.\\

\noindent
The compactification of the Euclidean AdS$_3$ Einstein gravity theory provides the nearly AdS$_2$/CFT$_1$ correspondence describing the relationship between the Euclidean Jackiw-Teitelboim (JT) gravity theory \cite{Jackiw:1984je} and the Sachdev-Ye-Kitaev (SYK) model. From the JT gravity theory, one derived the Euclidean Schwarzian theory on the boundary by integrating out the dilaton field \cite{Jensen:2016pah, Maldacena:2016upp}. The SYK model has all-to-all interaction from a four-Majorana fermion coupled term with a Gaussian random coupling constant. The SYK (or SYK$_4$) model also gives the Euclidean Schwarzian theory in the holographic limit, $N\gg \beta J\gg 1$. 
The $N$ is the number of Majorana fermion fields, $\beta$ is the inverse temperature, and $J$ is the coupling constant \cite{Engelsoy:2016xyb}. 
The Euclidean Schwarzian theory is not a conformal field theory. 
This theory is dual to the 1-dimensional Liouville theory \cite{Engelsoy:2016xyb, Bagrets:2016cdf}, which has given from the 2-dimensional Liouville theory through the dimensional reduction. 
The technology of the path integration in the Euclidean Schwarzian theory \cite{Stanford:2017thb} has been developed by using the Duistermaat-Heckman formula \cite{Duistermaat:1982vw}. 
The formula is the inverse Fourier transformation of a symplectic measure that can be exactly given by the stationary phase approximation \cite{Duistermaat:1982vw}. 
The formula was also generalized to the equivalent cohomology to enable doing integration, which is the same as summing overall fixed points of the Hamiltonian \cite{Atiyah:1984px}. Therefore, these results provide an interesting exact study going beyond the AdS/CFT correspondence. 
\\

\noindent
Because a derivation of the higher spin sector has not done so far. We want to extend the {\it SL(2)} study to the {\it SL(3)} theory for a study of the spin three \cite{Campoleoni:2010zq, Li:2015osa, Campoleoni:2017xyl, Narayan:2019ove}.
The central direction that we would like to address in this article is: {\it Could we obtain an integrable sector in Chern-Simons formulation?} In this paper, we show that the action of the Chern-Simons higher spin theory, after the dimensional reduction, is the action of {\it SL(3) Schwarzian theory}, which is not CFT. We also explicitly show that the Lorentzian SL(3) Schwarzian theory is dual to the integrable quantum mechanics, {\it SL(3) open Toda chain theory}. The first-order formulation of the JT gravity is BF theory \cite{Isler:1989hq}. The SL(3) Schwarzian theory has also been derived from the SL(3) BF theory \cite{Gonzalez:2018enk}. Hence  this gives a consistency between the dimensional reduction and BF theory \cite{Achucarro:1992mb}. Moreover, we discuss the application of open Toda chain theory from the perspective of quantum chaos.\\

\noindent
The dynamics of the SYK model provides the interesting Gaussian random matrix spectrum in the {\it spectral form factors} \cite{Dyer:2016pou, Cotler:2016fpe} numerically. One of the required conditions of the quantum chaos is the random eigenvectors \cite{Berry:1977zz}. 
This condition has been realized from the {\it dip-ramp-plateau} behavior of the spectral form factor at a high-temperature limit. 
The behavior first decreases, then increases, and then reaches an approximately constant in the time evolution \cite{Cotler:2016fpe}. This realization was justified from the SYK$_2$ model \cite{Lau:2018kpa}. Recently, one got an exact solution of the spectral form factor in the Gaussian unitary ensemble matrix model to confirm the {\it dip-ramp-plateau} behavior \cite{Okuyama:2018yep}. Since the SL(2) open Toda chain theory has infinite energy levels, it is similar to quantum field theory. It is still unclear whether our experience of quantum chaos can directly applies to quantum field theory. We choose the SL(2) open Toda chain theory, which has an infinite energy level, to do the numerical study for finding the evidence from the spectral form factor.
\\

\noindent
In this paper, we are mainly interested in the SL(3) Chern-Simons formulation because its effective action and the relevant analysis has not been derived from a correct way \cite{Cotler:2016fpe}. 
Our derivation gives the SL(3) Schwarzian theory at a low-energy limit, and it is valid at a quantum level. 
We can use the localization to obtain the partition function for the SL(3) Schwarzian theory and extract the spectral form factor from the partition function. 
Because the exact solutions between the SL(2) and SL(3) theories are similar for $\beta\neq 0$, we only show our numerical analysis for the SL(2) open Toda chain theory, which is dual to the SL(2) Schwarzian theory, to study the asymptotic behavior near $\beta=0$ on a lattice. 
The result does not contradict with the integrability.We show that the BF model, which is a gauge formulation of JT gravity theory, does not have a Gaussian random matrix spectrum. 
Our conclusion can also extend to the SL(3) Toda chain theory.

\section{Chern-Simons Higher Spin Theory and\\
 SL(3) Schwarzian Theory}
\label{sec:2}
\noindent
The action of the SL($M$) Chern-Simons formulation in the Lorentzian spacetime is given by \cite{Witten:1988hc}
\bea
S_{\mathrm{G}}&=&\frac{k}{2\pi}\int d^3x\ \epsilon^{tr\theta}\mathrm{Tr}\bigg(A_tF_{r\theta}-\frac{1}{2}\big(A_r\partial_tA_{\theta}-A_{\theta}\partial_tA_r\big)\bigg)
\nn\\
&&-\frac{k}{2\pi}\int d^3x\ \epsilon^{tr\theta}\mathrm{Tr}\bigg(\bar{A}_t\bar{F}_{r\theta}-\frac{1}{2}\big(\bar{A}_r\partial_t\bar{A}_{\theta}-\bar{A}_{\theta}\partial_t\bar{A}_r
\big)\bigg)
\nn\\
&&-\frac{k}{4\pi}\int dtd\theta\ \mathrm{Tr}(A_{\theta}^2)
\nn\\
&&-\frac{k}{4\pi}\int dtd\theta\ \mathrm{Tr}(\bar{A}_{\theta}^2),
\eea
in which the boundary conditions of the gauge fields $A$ and $\bar{A}$ are given by $A_t-A_\theta= 0$ and $\bar{A}_t+\bar{A}_\theta= 0$. The time direction is $t$, and two spatial directions are $r$ and $\theta$. Each bulk term, which lives in three dimensions, is equivalent to the action of the Chern-Simons theory up to a total derivative term. We assume that the gauge fields are given by $A_{\mu}\equiv A_{\mu}{}^aJ_a$ and $\bar{A}_{\mu}=\bar{A}_{\mu}{}^a\bar{J}_a$, in which the spacetime indices are labeled by $\mu=t, r, \theta$, and the Lie algebra indices are labeled by $a=1, 2, \cdots, M^2-1$. The algebra indices are raised or lowered by $\eta^{ab}\equiv\mathrm{diag}(-1,1,\cdots,1)$. The $J^a$ and $\bar{J}^a$ are the generators of the SL($M$) Lie algebras. The constant $k$ is $l/(4G_3)$ with $1/l^2\equiv-\Lambda$, where the cosmological constant is denoted by $\Lambda$, and 3-dimensional gravitational constant is denoted by $G_3$. We will choose the unit $\Lambda=-1$. The gauge fields are defined by the vielbeins $e_{\mu}{}^a$ and spin connections $\omega_{\mu}{}^a$ \cite{Witten:1988hc}: $A_{\mu}=J_a\big(e_{\mu}{}^a/l+\omega_{\mu}{}^a\big)$ and $\bar{A}_{\mu}=\bar{J}_a\big(e_{\mu}{}^a/l-\omega_{\mu}{}^a\big)$.
The path integral measure of the SL($M$) Chern-Simons formulation is $\int {\cal D}A{\cal D}\bar{A}$.\\

\noindent
When $M=2$, the theory is the spin-2 theory, which is the first-order formulation of Einstein gravity theory \cite{Witten:1988hc}. The boundary action of the spin-2 theory is just an SL(2) invariant double copy of the 2-dimensional Schwarzian theory \cite{Cotler:2018zff}. Ones could compactify the Euclidean time direction to obtain the Schwarzian theory \cite{Cotler:2018zff, Achucarro:1992mb}. \\

\noindent
In this paper, we focus on $M=3$. We are interested in the gauge fields in the Lorentzian theory, which provides the Lorentzian AdS$_3$ metric on the boundary ($r \rightarrow\infty$), namely the asymptotically AdS solution \cite{Campoleoni:2010zq}. 
We thus impose the following conditions on the gauge fields: $A_{r\to \infty}-A_{{\rm AdS}_3}={\cal O}(1)$ and $\bar{A}_{r\to \infty}-\bar{A}_{{\rm AdS}_3}={\cal O}(1)$,
which is known as the fall-off condition of the asymptotically AdS solution \cite{Campoleoni:2010zq,Campoleoni:2017xyl}.\\

\noindent
The equations of motion of Chern-Simons theory provide $F_{\mu\nu}^a=\bar{F}^a_{\mu\nu}=0$, where $F^a_{\mu\nu}$ and $\bar{F}^a_{\mu\nu}$ are the field strengths associated to the one-form gauge fields, $A$ and $\bar{A}$.
We integrate out ${A_t}$ and ${\bar{A}_t}$, which is equivalent to using the solutions of equations of motion, $A_I=g^{-1}\partial_Ig$ and $\bar{A}_I=\bar{g}^{-1}\partial_I\bar{g}$, to obtain the action written in terms of $g$ and $\bar{g}$. The spatial index is labeled by $I$. For $M=3$, the bulk terms of higher spin action cannot be just described by a lower dimension term, which is not the same as the situation in $M=2$. We have 16 independent variables to describe the $M=3$ theory, which is too complicated to analyze. For simplicity, we do the dimensional reduction \cite{Achucarro:1992mb} in the Euclidean time direction \cite{Cotler:2018zff} to study the $M=3$ theory. Because the Lorentzian and Euclidean Einstein gravity theories have the opposite overall signs in the actions, the action of the $M=3$ theory in the Euclidean time, which becomes $\theta$ in the boundary theory, is $S_{\mathrm{G}1}=(k/2)\int d\theta\ \mathrm{Tr}\big(g^{-1}(\partial_{\theta}g)g^{-1}(\partial_{\theta}g)\big)$ in the infrared limit.\\

\noindent
We use the Gauss decomposition to parameterize the SL(3) group elements
\bea
g_{\mathrm{SL(3)}(F, \lambda, \Psi)}&=&
\begin{pmatrix}
1&0&0
\\
F_1&1&0
\\
F_2&F_3&1
\end{pmatrix}
\begin{pmatrix}
\lambda_1&0&0
\\
0&\lambda_2&0
\\
0&0&\frac{1}{\lambda_1\lambda_2}
\end{pmatrix}
\begin{pmatrix}
1& \Psi_1&\Psi_2
\\
0&1&\Psi_3
\\
0&0&1
\end{pmatrix}.
\eea
The asymptotically AdS solution gives the following boundary conditions on the fields ($F$, $\lambda$, and $\Psi$): 
\bea
F_3&=&\frac{\partial_{\theta}F_2}{\partial_{\theta}F_1}; \qquad \lambda_2^3=\frac{\partial_{\theta}F_1}{\partial_{\theta}F_3}; \qquad \lambda^3_1=r^3\frac{1}{(\partial_{\theta} F_1)^2(\partial_{\theta}F_3)};
\nn\\
\Psi_1&=&\frac{\partial_{\theta}\lambda_1}{r\lambda_1}; \qquad \Psi_2=\frac{\partial_{\theta}^2\lambda_1}{r^2\lambda_1}; \qquad
\Psi_3
=
\frac{1}{r}\bigg(\frac{\partial_{\theta}\lambda_1}{\lambda_1}+\frac{\partial_{\theta}\lambda_2}{\lambda_2}\bigg).
\eea
Then we use these boundary conditions to obtain the action of the higher spin theory in the Euclidean time
\bea
&&S_{\mathrm{G}1}
\nn\\
&=&
-k\int d\theta\ \bigg\lbrack
\frac{\partial_{\theta}^3F_1}{\partial_{\theta}F_1}
+\frac{\partial_{\theta}^3F_3}{\partial_{\theta}F_3}
-\frac{4}{3}\bigg(\frac{\partial_{\theta}^2F_1}{\partial_{\theta}F_1}\bigg)^2
\nn\\
&&
-\frac{4}{3}\bigg(\frac{\partial_{\theta}^2F_3}{\partial_{\theta}F_3}\bigg)^2
-\frac{1}{3}\frac{\big(\partial_{\theta}^2F_1\big)\big(\partial_{\theta}^2F_3\big)}{\big(\partial_{\theta}F_1\big)\big(\partial_{\theta}F_3\big)}
\bigg\rbrack.
\eea
The measure of the SL(3) Schwarzian theory is given by 
\bea
\int dF_1\wedge dF_2\ \frac{1}{\big(\partial_{\theta}F_1\big)^2\bigg\lbrack\partial_{\theta}\bigg(\frac{\partial_{\theta}F_2}{\partial_{\theta}F_1}\bigg)\bigg\rbrack}.
\eea
The detail of the SL(3) measure is given in \ref{appa}.
This is invariant under the SL(3) transformation 
\bea
h_{\mathrm{SL(3)}}\cdot g_{\mathrm{SL(3)}}(F, \lambda, \Psi)=g_{\mathrm{SL(3)}} (\tilde{F}, \tilde{\lambda}, \tilde{\Psi}),
\eea
 where $h_{\mathrm{SL(3)}}$ is an SL(3) transformation and does not dependent on the fields. The transformed fields are denoted by $\tilde{F}$, $\tilde{\lambda}$, and $\tilde{\Psi}$. Moreover, the transformations of $F_1$ and $F_2$ reproduce the standard transformations in the SL(3) Schwarzian theory.

\section{SL(3) Open Toda Chain Theory}
\label{sec:4}
\noindent
We begin from the action in the Euclidean time 
\bea
&&S_{\mathrm{OTC1}}
\nn\\
&=&4k\int d\theta\ \bigg(\big(\partial_{\theta}\phi_1\big)^2
+\big(\partial_{\theta}\phi_2\big)^2
+\big(\partial_{\theta}\phi_1\big)\big(\partial_{\theta}\phi_2\big)
\nn\\
&&
+\Pi_{F_1}\big(\partial_{\theta}F_1-e^{2\phi_1-2\phi_2}\big)+\Pi_{F_2}\big(\partial_{\theta}F_2-e^{4\phi_2+2\phi_1}\big)\bigg).
\eea
The measure is 
$
\int (dF_1\wedge dF_2)(d\phi_1\wedge d\phi_2)(d\Pi_{F_1}\wedge d\Pi_{F_2})\
(1/\partial_{\theta}\phi_2)$.
If we integrate out the momenta, $\Pi_{F_1}$ and $\Pi_{F_2}$, the action goes back to the action $S_{\mathrm{G1}}$ up to a total derivative term and its measure. Now we integrate out the fields, $F_1$ and $F_2$, and choose $\Pi_{F_1}=-\lambda_1$ and $\Pi_{F_2}=-\lambda_2$, where $\lambda_1$ and $\lambda_2$ are arbitrary constants.
After doing the Wick rotation ($\theta\rightarrow i\theta$), introducing field $\phi_3=-(\phi_1+\phi_2)$, and doing the field redefinition: $\phi_1\rightarrow \phi_1-(1/6)\ln(\lambda_1^2\lambda_2)$, $\phi_2\rightarrow\phi_2-(1/6)\ln(\lambda_2/\lambda_1)$, and $\phi_3\rightarrow\phi_3+(1/6)\ln(\lambda_1\lambda_2^2)$, we obtain the action of the SL(3) open Toda chain theory 
\bea
S_{\mathrm{OTC}}=4k\int d\theta\ \bigg(
\frac{1}{2}\sum_{i_1=1}^{3}\big(\partial_{\theta}\phi_{i_1}\big)^2-\sum_{i_2=1}^{2}e^{2(\phi_{i_2}-\phi_{i_2+1})}\bigg).
\eea
The measure of the SL(3) open Toda Chain theory is
$\int d\phi_1\wedge d\phi_2\wedge d\phi_3\  \delta(\phi_1+\phi_2+\phi_3)/\partial_{\theta}\phi_2$.

\section{Spectral Form Factor in the SL(2) and SL(3) Open Toda Chain Theories}
\label{sec:5}
\noindent
The spectral form factor \cite{Dyer:2016pou, Cotler:2016fpe} is defined by $
g(t)\equiv\left|Z(\beta,t)\right|^2\big/\left|Z(\beta,0)\right|^2$, where $Z(\beta,t)\equiv\mathrm{Tr} \bigg( \exp\big(-(\beta-it)H\big) \bigg)$ is the un-normalized thermal average of operator $\exp(it H)$ and $H$ is the Hamiltonian of a system.\\

\noindent
The exact solution of the spectral form factor in the SL(2) open Toda chain theory $g(t)=\big(\beta^3/(\beta^2+t^2)^{3/2}\big)\cdot\exp\bigg(-\pi^2 t^2/\big(\beta(\beta^2+t^2)\big)\bigg)$ \cite{Cotler:2016fpe} can be obtained by the partition function of the SL(2) Schwarzian theory
$Z_{\mathrm{S}2}(\beta, 0)\sim \exp\big(\pi^2/(2\beta)\big)/\beta^{3/2}$ through the localization \cite{Stanford:2017thb}, in which we choose $k=1/2$, and its analytical continuation on time. Therefore, we can find that the spectral form factor only has the dip except for $\beta=0$. When $\beta=0$, we have $Z_{\mathrm{S2}}(\beta=0, t)=1/t^3$. This gives a divergent at $t=0$. Hence the spectral form factor is divergent for the normalization at $t=0$. This should be due to the issue of an infinite summation. However, this should not be problematic by using the lattice model. Hence we discretize the SL(2) open Toda chain theory to study the numerical solution of spectral form factor at $\beta=0$. Due to $Z_{\mathrm{S2}}(\beta=0, t)$, we expect that a lattice study should give a decay as the $1/t^3$ if the normalization issue can be ignored. 
\\

\noindent
The Hamiltonian of the SL(2) open Toda chain theory is $H_{\mathrm{OTC}}=p^2+e^{4\phi}$,
where $p\equiv-i\partial/\partial\phi$. Here we choose $k=1/4$ because we hope that the pre-factor is 1 for convenience. In our numerical study, we put the theory on a lattice defined as the following: $-L\le\phi_{j}<L$, $\phi_{1}=-L$, $\phi_{j+1}\equiv\phi_{j}+a$, $\phi_{0}\equiv\phi_{n}$, $\phi_{n+1}\equiv \phi_{1}$, and $2L=n\cdot a$. The lattice index is labeled by $j=1, 2, \cdots, n$.
The lattice spacing is denoted by $a$, and the lattice size is denoted by $L$. The kinetic term on a lattice is written as $
p_l^2\psi_j\equiv-(\psi_{j+1}-2\psi_{j}+\psi_{j-1})/a^2$, where $p_l$ is the lattice canonical momentum, and $\psi_j$ is the lattice  eigenfunction. \\

\begin{figure}[h]
\begin{centering}
\includegraphics[width=0.3\textwidth]{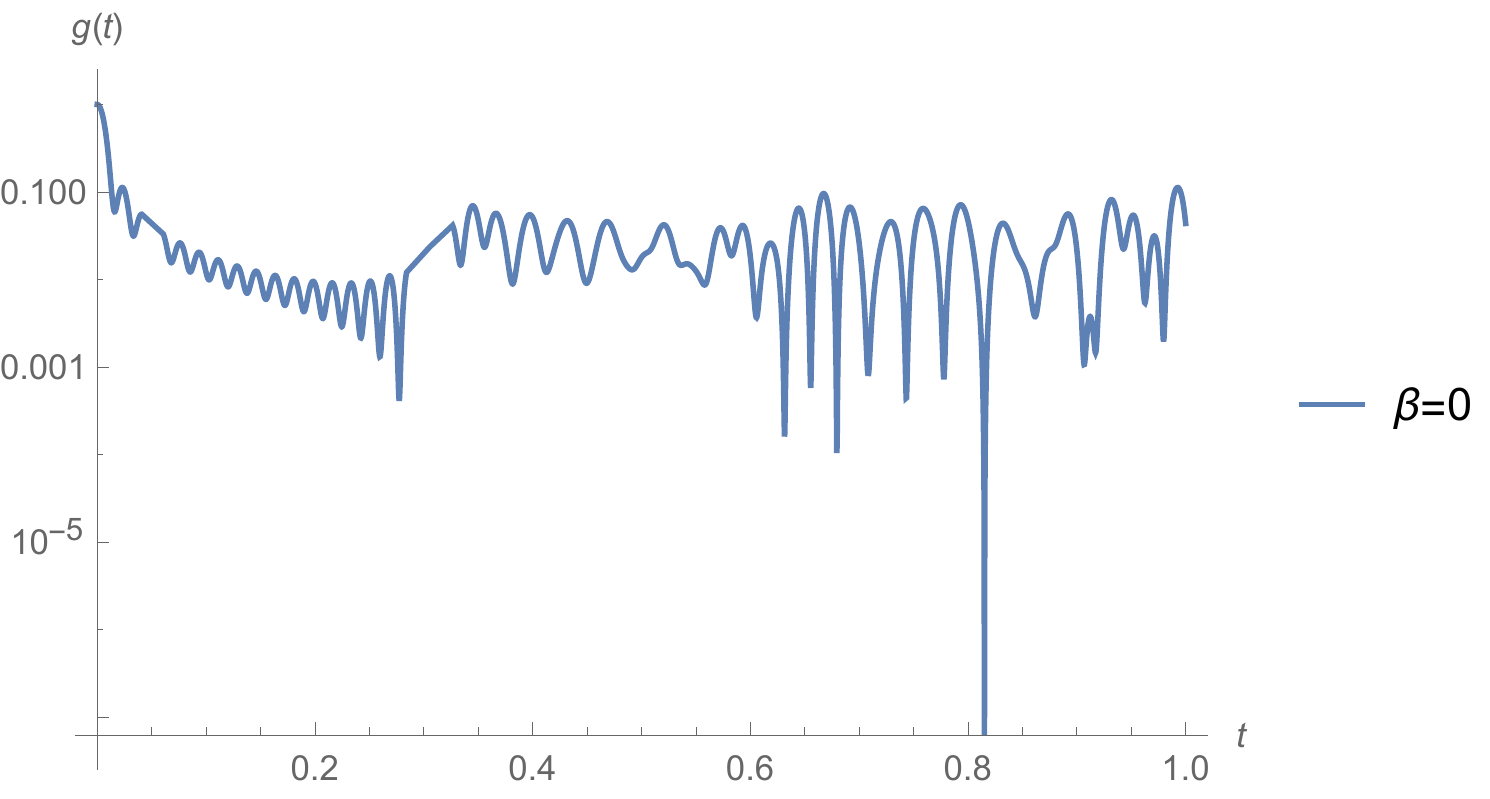}
\hfill
\includegraphics[width=0.3\textwidth]{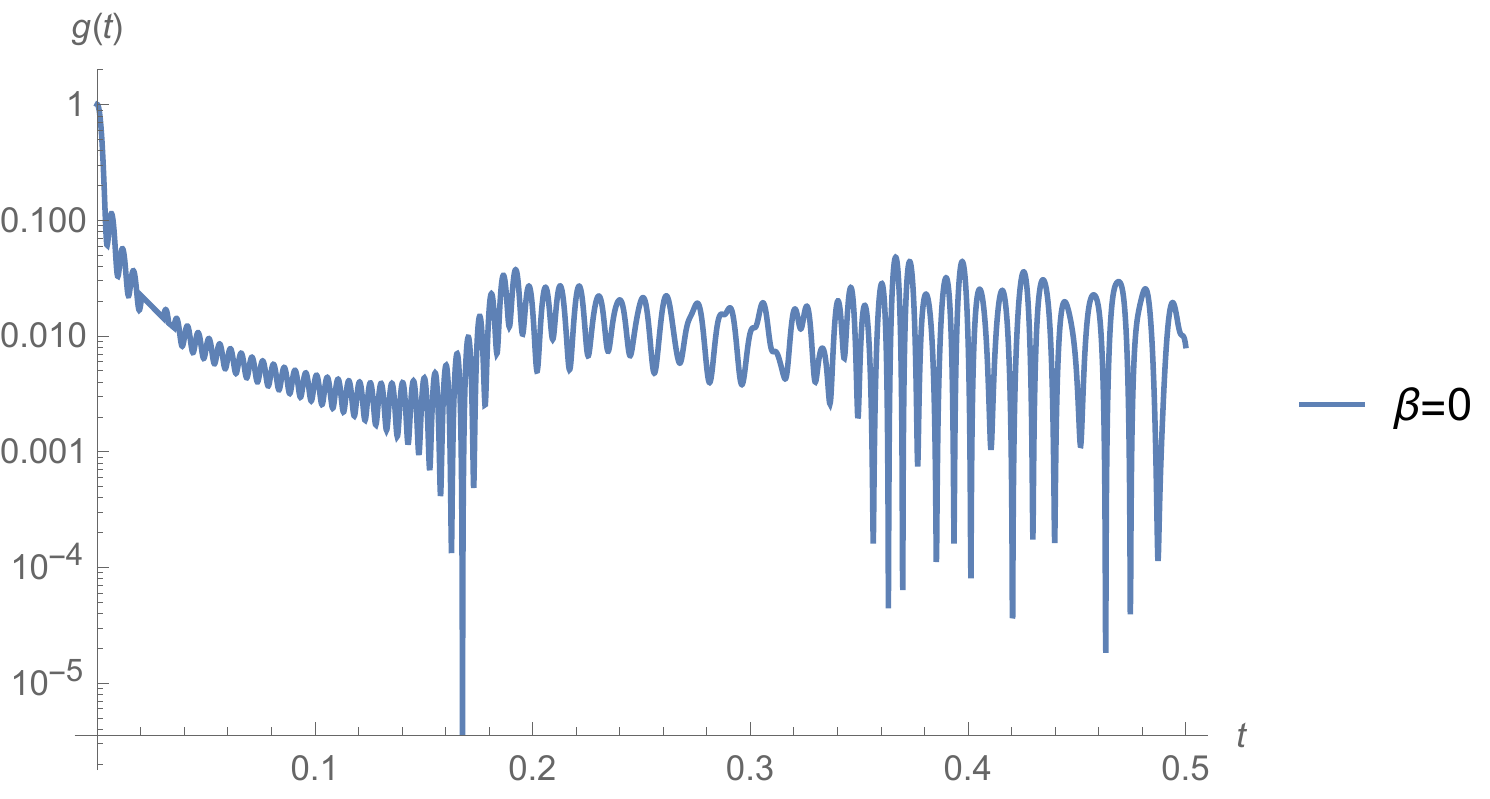}
\hfill
\includegraphics[width=0.3\textwidth]{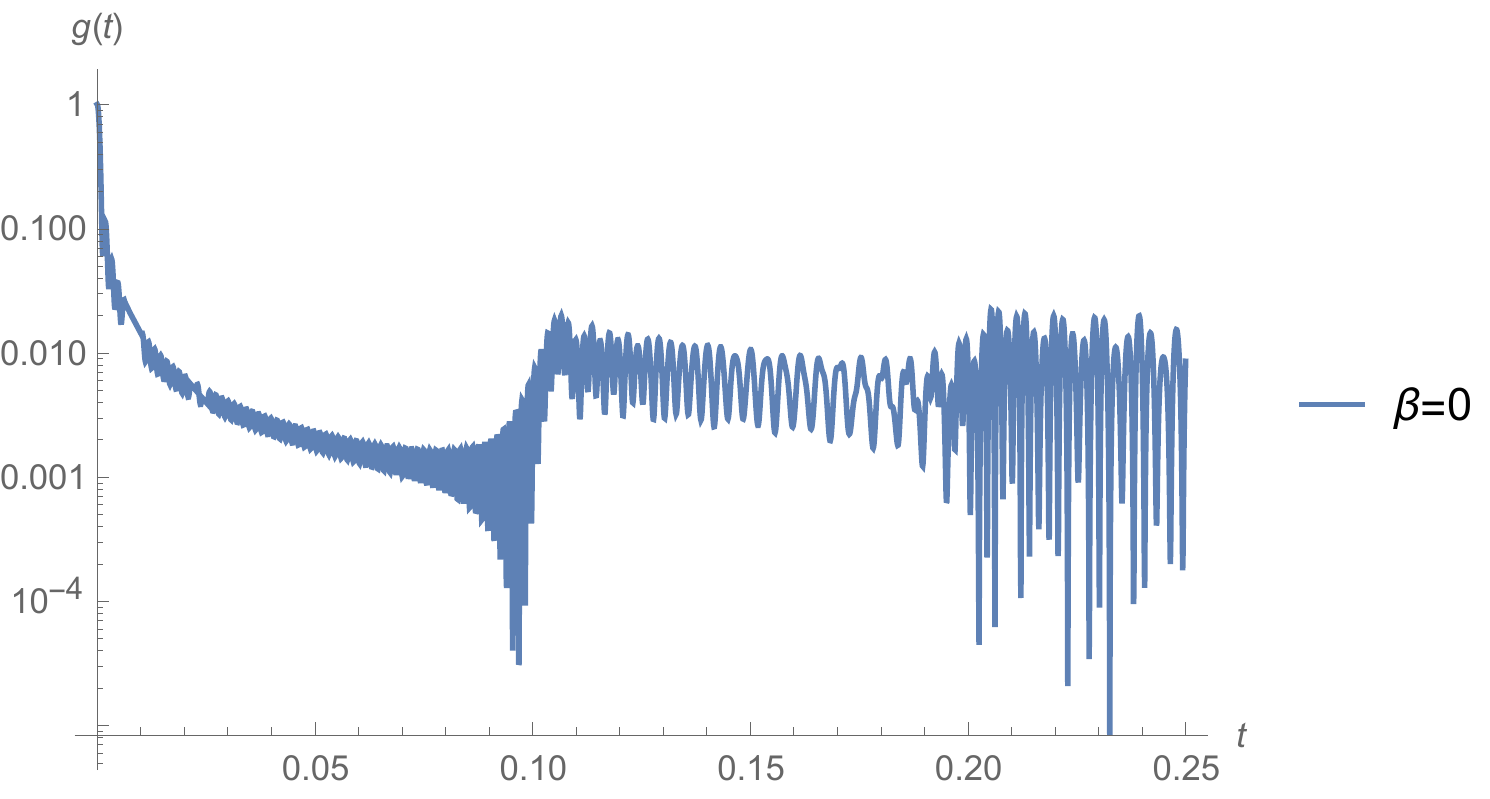}
\par\end{centering}
\caption{We fix the inverse temperature $\beta=0$ while choosing the lattice size $L=4$ and the number of lattice points $n=1024$. The results of the spectral form factor $g(t)$ are for 32, 64, and 128 low-lying energy modes on the left, middle, and right figures respectively. Although $g(t)$ shows the dip-ramp-plateau behavior from the low-lying modes, the ramp time changes from the number of the low-lying modes. Hence the dip-ramp-plateau behavior should be caused by the lattice artifact. The result is consistent with the integrability. The value of $g(t)$ is written as the log scale.
}
\label{spm4.pdf}
\end{figure}

\noindent
Since the theory puts in a box in our numerical study, the only realizable energy modes are the ones that are larger than the potential.  To provide a relizable result, we first fix the number of the low-lying modes and check the lattice spacing effects with a given lattice size in the spectral form factor. We fix the inverse temperature $\beta=0$ and choose the lattice size $L=4$ and the number of lattice points $n=1024$. In Fig. \ref{spm4.pdf}, we plot the spectral form factor to show the dip-ramp-plateau behavior for 32, 64, and 128 low-lying modes. Because the ramp time is roughly divided by two when the number of the low-lying modes doubles, the ramp time should approach zero under the continuum and the infinite size limits. Therefore, dip-ramp-plateau behavior in Fig. \ref{spm4.pdf} should be due to the lattice artifact. The relation between the BF theory and the SL(2) Toda chain theory is exact \cite{Achucarro:1992mb}. Hence spectrum of the BF theory should be integrable.
\\

\noindent
 Hence these results do not contradict the fact of integrability. Our numerical study possibly not be accurate enough to reach the quantitative level. The physical interpretation and qualitative behaviors should not be modified from the lattice size effect.
 \\
 
 \noindent
\\

\begin{figure}[h]
\begin{centering}
\includegraphics[width=0.49\textwidth]{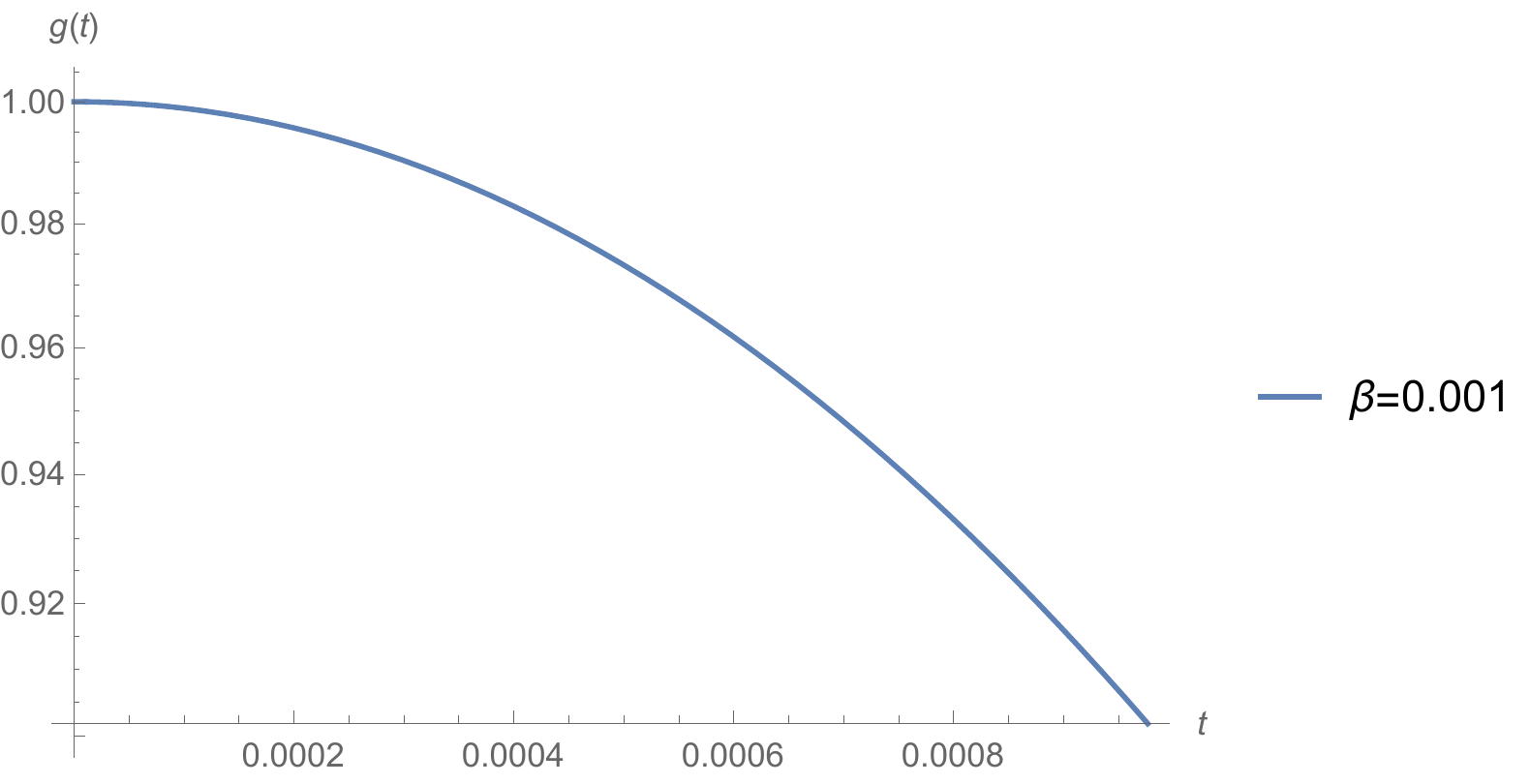}
\hfill
\includegraphics[width=0.49\textwidth]{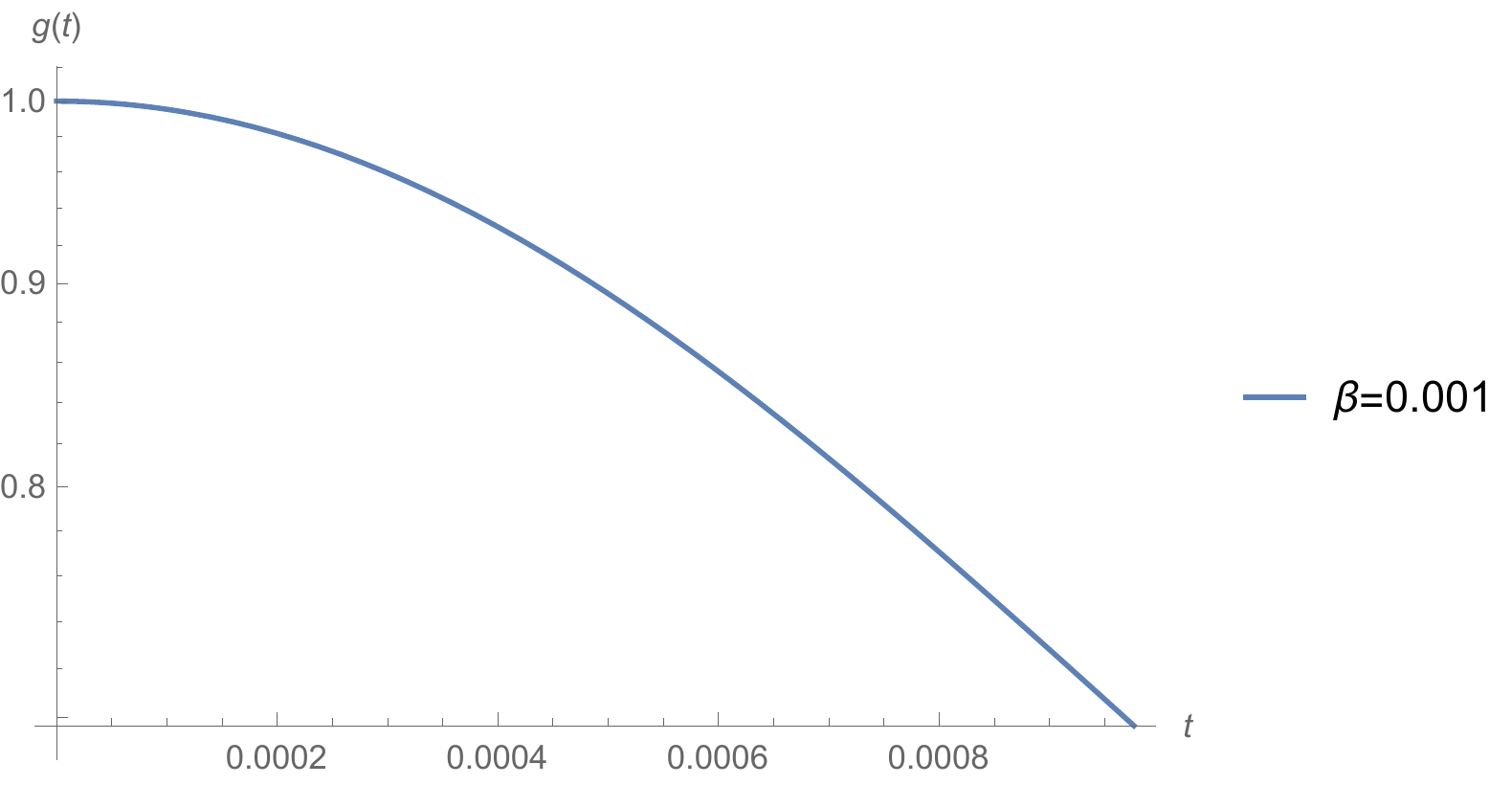}
\par\end{centering}
\caption{We fix the inverse temperature $\beta=0.001$ while choosing the lattice size $L=4$ and the number of lattice points $n=1024$. The results of the spectral form factor $g(t)$ are for 64 and 128 low-lying energy modes on the left and right figures respectively. Increasing the number of low-lying modes tends to have a more decay. The value of $g(t)$ is written as the log scale.
}
\label{beta_0.001.pdf}
\end{figure}

\begin{figure}[h]
\begin{centering}
\includegraphics[width=0.49\textwidth]{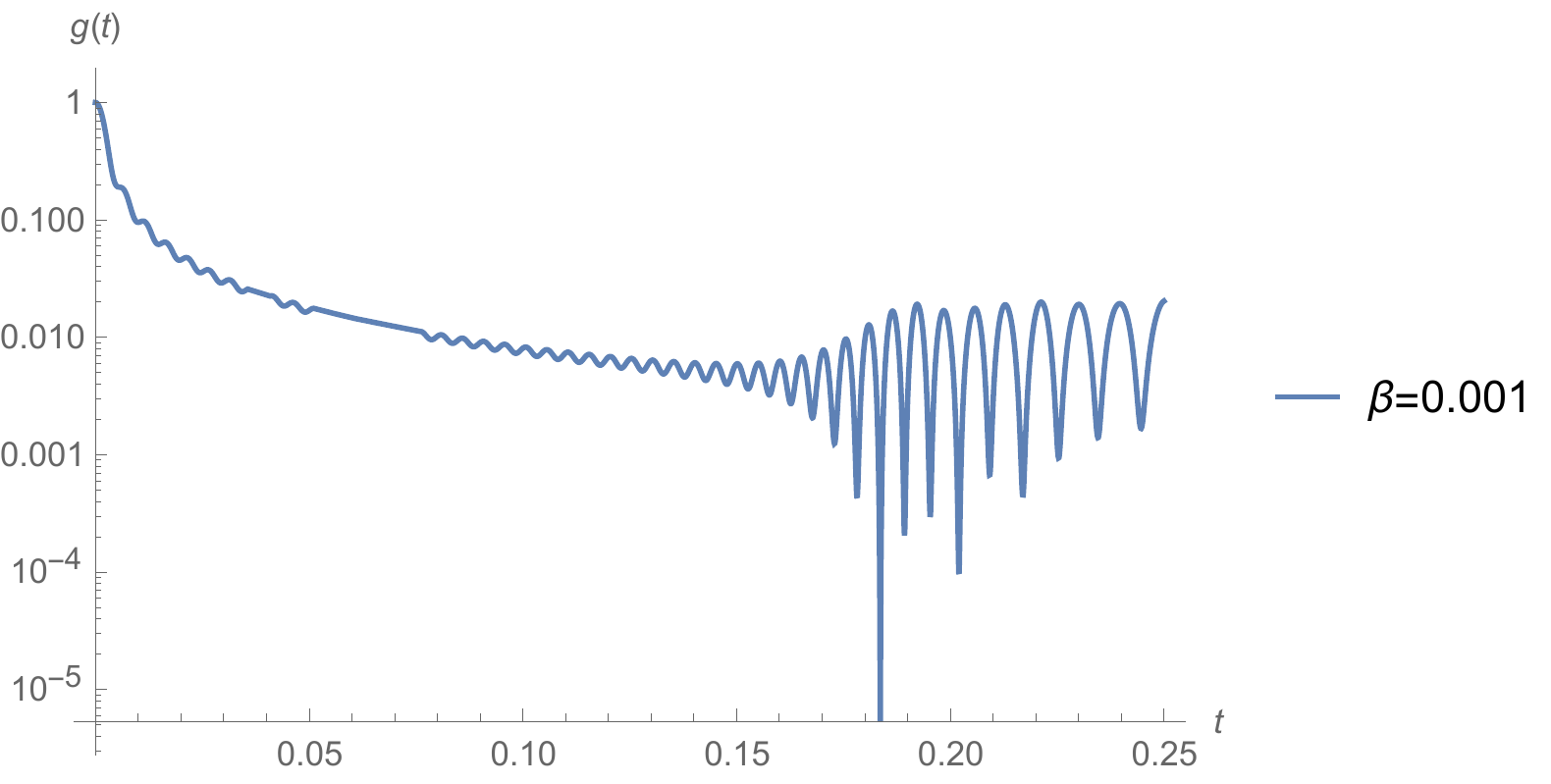}
\hfill
\includegraphics[width=0.49\textwidth]{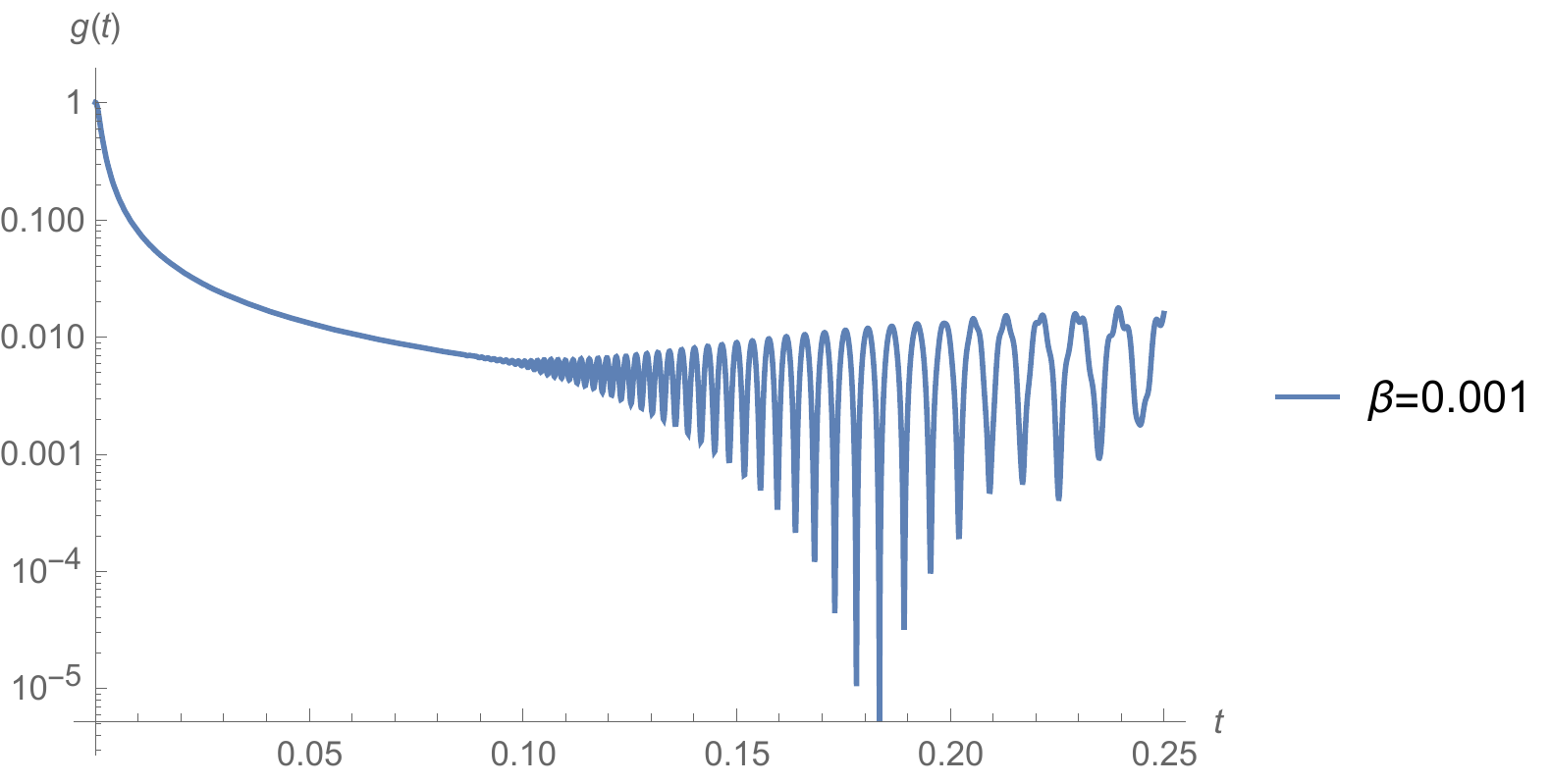}
\par\end{centering}
\caption{We fix the inverse temperature $\beta=0.001$ while choosing the lattice size $L=4$ and the number of lattice points $n=1024$. The results of the spectral form factor $g(t)$ are for 64 and 128 low-lying energy modes on the left and right figures respectively. The figures do not show the dip-ramp-plateau behavior. The value of $g(t)$ is written as the log scale.
}
\label{Nbeta_0.001.pdf}
\end{figure}

\noindent
Because the numerical study is not accurate at $\beta=0$, we analyzed the numerical study at $\beta=0.001$. At $\beta=0.001$, choosing the $L=4$ and $n=1024$ is also enough to obtain a qualitative result to the exact solution, $g(t)=\big(\beta^3/(\beta^2+t^2)^{3/2}\big)\cdot\exp\bigg(-\pi^2 t^2/\big(2\beta(\beta^2+t^2)\big)\bigg)$. 
The exact solution exhibits a fast decay at $\beta=0.001$ even for a short time. 
The numerical solution does not show such a decay, but we find that more numbers of low-lying modes give faster decay in Fig. \ref{beta_0.001.pdf}. 
If we can do a large number of summation for low-lying modes, we should obtain the exact solution. 
Our numerical result also shows no dip-ramp-plateau behavior as the exact solution in Fig. \ref{Nbeta_0.001.pdf}. 
The oscillating behavior in Fig. \ref{Nbeta_0.001.pdf} should be due to a lattice artifact because the exact solution does not have such dynamics.
\\

 \noindent
\\

\noindent
In the final, we show the exact solution of the spectral form factor for the SL(3) theory. From the localization, the partition function of the SL(3) theory is $Z_{\mathrm{S}3}(\beta, 0)\sim \exp\big(2\pi^2/\beta\big)/\beta^{4}$, which is for $k=1/2$. The classical term gives the exponentiation function. The $1/\beta^4$ is contributed by the one-loop term, and its power due to the eight zero modes \cite{Stanford:2017thb, Achucarro:1992mb}. Then it is easy to show the spectral form factor $g(t)=\big(\beta^8/(\beta^2+t^2)^4\big)\cdot\exp\bigg(-4\pi^2 t^2/\big(\beta(\beta^2+t^2)\big)\bigg)$. Hence the behavior of the spectral form factor for the SL(3) theory also has a similar form to the SL(2) theory for $\beta\neq 0$. We should expect that the asymptotic behaviors are similar to the SL(2) case at $\beta=0$ on a lattice. Hence the SL(3) theory should not have a Gaussian random matrix spectrum.

\section{Outlook}
\label{sec:6} 
\noindent 
In this paper, we have shown that the dimensional reduction of the boundary action in SL(3) Chern-Simons theory is the SL(3) reparametrization invariant Schwarzian theory. A duality between the Lorentzian SL(3) Schwarzian theory and the SL(3) open Toda chain theory was derived. 
Because the old derivation of Chern-Simons formulation is not correct at a quantum level \cite{Coussaert:1995zp}, we adopted a correct derivation \cite{Cotler:2018zff} to the SL(3) case, which is a continuation of the SL(2) case \cite{Cotler:2018zff}. 
Because the SL(3) Chern-Simons formulation is connected to the SL(3) Schwarzian theory or open Toda chain theory, similar to the SL(2) case, we can obtain an exact solution in the spectral form factor as in the SL(2) case. 
Moreover, the behavior of the exact solutions between the SL(2) and SL(3) cases are similar. Therefore, we only analyzed the SL(2) case without showing the analysis of SL(3) because no new physical results will be obtained.
Because the BF formulation is dual to the SL(2) Toda chain theory \cite{Engelsoy:2016xyb, Bagrets:2016cdf}, we justified the existence of a Gaussian random matrix spectrum and duality. A suitable quantity for diagnosing chaos should be valid for any duality. Because the dip-ramp-plateau behavior in the spectral form factor is related to the randomness \cite{Lau:2018kpa}, this should be a condition of the irregular motion under the classical limit \cite{Berry:1977zz}. The irregular motion in the classical chaos in the interval is a purely topological condition without any affecting by any mapping. Moreover, our result in the SL(2) Toda chain theory should demonstrate a useful application in the higher spin theory or different limit of string theory because the open Toda chain theories have a similar form (like the exact solution for the $M=2$ and $M=3$ in the spectral form factor) and are all integrable.
\\

\noindent 
Although the open Toda chain theory is a quantum mechanical system, it has infinite energy levels. Quantum chaos in quantum field theory still has ambiguities in the definition.  Hence the open Toda chain theory is also a good toy model to explore quantum chaos in quantum field theory. One can also do an exact diagonalization to calculate the correlation function in the numerical study. Therefore, the interesting holographic principle of the higher spin model can give from our generalization.
\\

\noindent 
Because the SL(3) BF theory and SL(3) Chern-Simons higher spin theory after the dimensional reduction have the same partition function (SL(3) Schwarzian theory) \cite{Gonzalez:2018enk}, two theories should be equivalent. 
One needs to do a non-trivial field redefinition to obtain the SL(2) BF theory from the SL(2) Chern-Simons formulation after the dimensional reduction \cite{Achucarro:1992mb}. 
Hence we expect a direct derivation from a similar field redefinition to explore a more clear study.

\section*{Acknowledgments}
\noindent 
We would like to thank Jordan Cotler, Kristan Jensen, Pak Hang Chris Lau, Shiraz Minwalla, Jeff Murugan, and Tadashi Takayanagi for their useful discussions. 
Chen-Te Ma would like to thank Nan-Peng Ma for his encouragement. 
\\

\noindent 
 Chen-Te Ma was supported by the Post-Doctoral International Exchange Program; 
 China Postdoctoral Science Foundation, Postdoctoral General Funding: Second Class (Grant No. 2019M652926); 
 Science and Technology Program of Guangzhou (Grant No. 2019050001). 
Hongfei Shu was supported by the JSPS Research Fellowship 17J07135 for Young Scientists from Japan Society for the Promotion of Science (JSPS); 
the grant ``Exact Resultsin Gauge and String Theories'' from the Knut and Alice Wallenberg foundation. 
\\

\noindent
We would like to thank National Tsing Hua University and Jinan University.
\\

\noindent
Discussions during the workshops, ``NCTS Annual Theory Meeting 2018: Particles, Cosmology and Strings'' and ``Jinan University Gravitational Frontier Seminar'', were useful to complete this work.

\appendix
\section{SL(3) Measure}
\label{appa}
\noindent
The SL(3) transformation is  
\bea
g_{\mathrm{SL(3)}}
=
\begin{pmatrix}
\lambda_1& \lambda_1\Psi_1&\lambda_1\Psi_2
\\
\lambda_1F_1&\lambda_1F_1\Psi_1+\lambda_2&\lambda_1F_1\Psi_2+\lambda_2\Psi_3
\\
\lambda_1F_2&\lambda_1F_2\Psi_1+\lambda_2F_3&\lambda_1F_2\Psi_2+\lambda_2F_3\Psi_3+\frac{1}{\lambda_1\lambda_2}
\end{pmatrix}.
\eea
\\


\noindent
Now we compute the derivative of the SL(3) transformation:
\bea
\frac{\partial g_{\mathrm{SL(3)}}}{\partial\lambda_1}&=&
\begin{pmatrix}
1&\Psi_1&\Psi_2
\\
F_1&F_1\Psi_1&F_1\Psi_2
\\
F_2&F_2\Psi_1&F_2\Psi_2-\frac{1}{\lambda_1^2\lambda_2}
\end{pmatrix};
\nn\\
\frac{\partial g_{\mathrm{SL(3)}}}{\partial\lambda_2}&=&
\begin{pmatrix}
0&0&0
\\
0&1&\Psi_3
\\
0&F_3&F_3\Psi_3-\frac{1}{\lambda_1\lambda_2^2}
\end{pmatrix};
\nn\\
\frac{\partial g_{\mathrm{SL(3)}}}{\partial F_1}&=&
\begin{pmatrix}
0&0&0
\\
\lambda_1&\lambda_1\Psi_1&\lambda_1\Psi_2
\\
0&0&0
\end{pmatrix}; 
\nn\\
\frac{\partial g_{\mathrm{SL(3)}}}{\partial F_2}&=&
\begin{pmatrix}
0&0&0
\\
0&0&0
\\
\lambda_1&\lambda_1\Psi_1&\lambda_1\Psi_2
\end{pmatrix}; 
\nn\\
\frac{\partial g_{\mathrm{SL(3)}}}{\partial F_3}&=&
\begin{pmatrix}
0&0&0
\\
0&0&0
\\
0&\lambda_2&\lambda_2\Psi_3
\end{pmatrix};
\nn\\
\frac{\partial g_{\mathrm{SL(3)}}}{\partial\Psi_1}&=&
\begin{pmatrix}
0&\lambda_1&0
\\
0&\lambda_1F_1&0
\\
0&\lambda_1F_2&0
\end{pmatrix}; 
\nn\\
\frac{\partial g_{\mathrm{SL(3)}}}{\partial\Psi_2}&=&
\begin{pmatrix}
0&0&\lambda_1
\\
0&0&\lambda_1F_1
\\
0&0&\lambda_1F_2
\end{pmatrix};
\nn\\
\frac{\partial g_{\mathrm{SL(3)}}}{\partial\Psi_3}&=&
\begin{pmatrix}
0&0&0
\\
0&0&\lambda_2
\\
0&0&\lambda_2F_3
\end{pmatrix}.
\eea
Then we can obtain the followings:
\bea
g^{-1}_{\mathrm{SL(3)}}\frac{\partial g_{\mathrm{SL(3)}}}{\partial\lambda_1}=
\begin{pmatrix}
\frac{1}{\lambda_1}& \frac{\Psi_1}{\lambda_1}&\frac{2\Psi_2-\Psi_1\Psi_3}{\lambda_1}
\\
0&0&\frac{\Psi_3}{\lambda_1}
\\
0&0&-\frac{1}{\lambda_1}
\end{pmatrix}; \qquad
g^{-1}_{\mathrm{SL(3)}}\frac{\partial g_{\mathrm{SL(3)}}}{\partial\lambda_2}=
\begin{pmatrix}
0& -\frac{\Psi_1}{\lambda_2}&\frac{\Psi_2-2\Psi_1\Psi_3}{\lambda_2}
\\
0&\frac{1}{\lambda_2}&2\frac{\Psi_3}{\lambda_2}
\\
0&0&-\frac{1}{\lambda_2}
\end{pmatrix};
\nn
\eea
\bea
&&g^{-1}_{\mathrm{SL(3)}}\frac{\partial g_{\mathrm{SL(3)}}}{\partial F_1}
\nn\\
&=&
\begin{pmatrix}
W_1
& W_1\Psi_1
&W_1\Psi_2
\\
\frac{\lambda_1}{\lambda_2}+\lambda_1^2\lambda_2F_3\Psi_3
&\frac{\lambda_1}{\lambda_2}\Psi_1+\lambda_1^2\lambda_2F_3\Psi_1\Psi_3
&\frac{\lambda_1}{\lambda_2}\Psi_2+\lambda_1^2\lambda_2F_3\Psi_2\Psi_3
\\
-\lambda_1^2\lambda_2F_3&-\lambda_1^2\lambda_2F_3\Psi_1&-\lambda_1^2\lambda_2F_3\Psi_2
\end{pmatrix};
\nn\\
&&g^{-1}_{\mathrm{SL(3)}}\frac{\partial g_{\mathrm{SL(3)}}}{\partial F_2}
\nn\\
&=&
\begin{pmatrix}
\lambda_1^2\lambda_2(\Psi_1\Psi_3-\Psi_2)
& \lambda_1^2\lambda_2(\Psi_1\Psi_3-\Psi_2)\Psi_1
&\lambda_1^2\lambda_2(\Psi_1\Psi_3-\Psi_2)\Psi_2
\\
-\lambda_1^2\lambda_2\Psi_3
&-\lambda_1^2\lambda_2\Psi_1\Psi_3
&-\lambda_1^2\lambda_2\Psi_2\Psi_3
\\
\lambda_1^2\lambda_2&\lambda_1^2\lambda_2\Psi_1&\lambda_1^2\lambda_2\Psi_2
\end{pmatrix};
\nn\\
&&g^{-1}_{\mathrm{SL(3)}}\frac{\partial g_{\mathrm{SL(3)}}}{\partial F_3}
\nn\\
&=&
\begin{pmatrix}
0
& \lambda_1\lambda_2^2(\Psi_1\Psi_3-\Psi_2)
&\lambda_1\lambda_2^2(\Psi_1\Psi_3-\Psi_2)\Psi_3
\\
0
&-\lambda_1\lambda_2^2\Psi_3
&-\lambda_1\lambda_2^2\Psi_3^2
\\
0&\lambda_1\lambda_2^2&\lambda_1\lambda_2^2\Psi_3
\end{pmatrix};
\nn\\
&&g^{-1}_{\mathrm{SL(3)}}\frac{\partial g_{\mathrm{SL(3)}}}{\partial \Psi_1}
\nn\\
&=&
\begin{pmatrix}
0
&1
&0
\\
0
&0
&0
\\
0&0&0
\end{pmatrix};
\nn\\
&&g^{-1}_{\mathrm{SL(3)}}\frac{\partial g_{\mathrm{SL(3)}}}{\partial \Psi_2}
\nn\\
&=&
\begin{pmatrix}
0
&0
&1
\\
0
&0
&0
\\
0&0&0
\end{pmatrix};
\nn\\
&&g^{-1}_{\mathrm{SL(3)}}\frac{\partial g_{\mathrm{SL(3)}}}{\partial \Psi_3}
\nn\\
&=&
\begin{pmatrix}
0
&0
&-\Psi_1
\\
0
&0
&1
\\
0&0&0
\end{pmatrix},
\eea
where
\bea
W_1\equiv-\frac{\lambda_1}{\lambda_2}\Psi_1-\lambda_1^2\lambda_2F_3(\Psi_1\Psi_3-\Psi_2).
\eea
Therefore, we obtain the followings:
\bea
G_{\lambda_1\lambda_1}&=&\frac{2}{\lambda_1^2}; \qquad 
G_{\lambda_1\lambda_2}=G_{\lambda_2\lambda_1}=\frac{1}{\lambda_1\lambda_2}; \qquad
G_{\lambda_2\lambda_2}=\frac{2}{\lambda_2^2};
\nn\\
G_{F_1\Psi_1}&=&G_{\Psi_1F_1}=\frac{\lambda_1}{\lambda_2}+\lambda_1^2\lambda_2F_3\Psi_3; \qquad G_{F_1\Psi_2}=G_{\Psi_2F_1}=-\lambda_1^2\lambda_2F_3;
\nn\\
G_{F_2\Psi_1}&=&G_{\Psi_1F_2}=-\lambda_1^2\lambda_2\Psi_3; \qquad G_{F_2\Psi_2}=G_{\Psi_2F_2}=\lambda_1^2\lambda_2; 
\nn\\
G_{F_3\Psi_3}&=&G_{\Psi_3F_3}=\lambda_1\lambda_2^2;
\eea
\bea
&&G_{\lambda_1F_1}=G_{F_1\lambda_1}=G_{\lambda_1F_2}=G_{F_2\lambda_1}=G_{\lambda_1F_3}=G_{F_3\lambda_1}
\nn\\
&&=G_{\lambda_1\Psi_1}=G_{\Psi_1\lambda_1}=G_{\lambda_1\Psi_2}=G_{\Psi_2\lambda_1}
=G_{\lambda_1\Psi_3}=G_{\Psi_3\lambda_1}
\nn\\
&&=G_{\lambda_2F_1}=G_{F_1\lambda_2}=G_{\lambda_2F_2}=G_{F_2\lambda_2}=G_{\lambda_2F_3}=G_{F_3\lambda_2}
\nn\\
&&=G_{\lambda_2\Psi_1}=G_{\Psi_1\lambda_2}=G_{\lambda_2\Psi_2}=G_{\Psi_2\lambda_2}=G_{\lambda_2\Psi_3}
=G_{\Psi_3\lambda_2}
\nn\\
&&=G_{F_1F_1}=G_{F_1F_2}=G_{F_2F_1}=G_{F_1F_3}=G_{F_3F_1}=G_{F_1\Psi_3}=G_{\Psi_3F_1}
\nn\\
&&=G_{F_2F_2}=G_{F_2F_3}=G_{F_3F_2}=G_{F_2\Psi_3}=G_{\Psi_3F_2}
\nn\\
&&=G_{F_3F_3}=G_{F_3\Psi_1}=G_{\Psi_1F_3}
=G_{F_3\Psi_2}=G_{\Psi_2F_3}
\nn\\
&&=G_{\Psi_1\Psi_1}=G_{\Psi_1\Psi_2}=G_{\Psi_2\Psi_1}=G_{\Psi_1\Psi_3}=G_{\Psi_3\Psi_1}
\nn\\
&&=G_{\Psi_2\Psi_2}
=G_{\Psi_2\Psi_3}=G_{\Psi_3\Psi_2}=G_{\Psi_3\Psi_3}
=0.
\eea
Hence the determinant of the matrix $G$ gives $-3\lambda_1^6\lambda_2^2$. Therefore, the measure is 
\bea
\int d\lambda_1\wedge d\lambda_2\wedge dF_1\wedge dF_2\wedge dF_3\wedge d\Psi_1\wedge d\Psi_2 \wedge d\Psi_3\ \lambda_1^3\lambda_2.
\eea
\\

\noindent
Now we impose the constraints of the boundary into the measure:
\bea
&&\int d\lambda_1 \wedge d\lambda_2 \wedge dF_1\wedge dF_2\wedge dF_3\wedge d\Psi_1\wedge d\Psi_2\wedge d\Psi_3\ \lambda_1^3\lambda_2
\nn\\ 
&&\times\delta\bigg(F_3-\frac{\partial_{\theta}F_2}{\partial_{\theta}F_1}\bigg)\delta\bigg(\lambda_1^3\big(\partial_{\theta}F_1\big)^2-\frac{r^3}{\partial_{\theta}F_3}\bigg)
\delta\bigg(\frac{\lambda_2^3\partial_{\theta}F_3}{\partial_{\theta}F_1}-1\bigg)
\nn\\
&&\times\delta\bigg(r\Psi_1-\frac{\partial_{\theta}\lambda_1}{\lambda_1}\bigg)
\delta\bigg(\Psi_2-\frac{\partial_{\theta}^2\lambda_1}{r^2\lambda_1}\bigg)
\delta\Bigg(\Psi_3-\frac{1}{r}\bigg(\frac{\partial_{\theta}\lambda_1}{\lambda_1}+\frac{\partial_{\theta}\lambda_2}{\lambda_2}\bigg)\Bigg)
\nn\\
&=&-\int d\lambda_1\wedge d\lambda_2\wedge dF_1\wedge dF_2\wedge dF_3\ \frac{\lambda_1^3\lambda_2}{r}
\nn\\ 
&&\times\delta\bigg(F_3-\frac{\partial_{\theta}F_2}{\partial_{\theta}F_1}\bigg)\delta\bigg(\lambda_1^3\big(\partial_{\theta}F_1\big)^2-\frac{r^3}{\partial_{\theta}F_3}\bigg)
\delta\bigg(\frac{\lambda_2^3\partial_{\theta}F_3}{\partial_{\theta}F_1}-1\bigg)
\nn\\
&\sim&\int dF_1\wedge dF_2\wedge dF_3\wedge  d\tilde{\lambda}_1\wedge d\tilde{\lambda}_2\ 
\bigg(\frac{\tilde{\lambda}_1}{\tilde{\lambda}_2}\bigg)^{\frac{1}{3}} 
\frac{1}{\big(\partial_{\theta}F_1\big)^2\big(\partial_{\theta} F_3\big)^{\frac{2}{3}}}\frac{1}{r}
\nn\\
&&\times \delta\bigg(F_3-\frac{\partial_{\theta}F_2}{\partial_{\theta}F_1}\bigg)
\delta\bigg(\tilde{\lambda}_1-\frac{r^3}{\partial_{\theta}F_3}\bigg)
\delta\bigg(\tilde{\lambda}_2-1\bigg)
\nn\\
&=&\int dF_1\wedge dF_2\wedge dF_3\wedge d\tilde{\lambda}_1\ 
\nn\\
&&\times\frac{\tilde{\lambda}_1^{\frac{1}{3}}}{\big(\partial_{\theta}F_1\big)^2\big(\partial_{\theta}F_3\big)^{\frac{2}{3}}}\frac{1}{r}
 \delta\bigg(F_3-\frac{\partial_{\theta}F_2}{\partial_{\theta}F_1}\bigg)
\delta\bigg(\tilde{\lambda}_1-\frac{r^3}{\partial_{\theta}F_3}\bigg)
\nn\\
&=&-\int dF_1\wedge dF_2\wedge dF_3\ \frac{1}{\big(\partial_{\theta}F_1\big)^2\big(\partial_{\theta}F_3\big)} \delta\bigg(F_3-\frac{\partial_{\theta}F_2}{\partial_{\theta}F_1}\bigg)
\nn\\
&=&-\int dF_1\wedge dF_2\ \frac{1}{\big(\partial_{\theta}F_1\big)^2\bigg\lbrack\partial_{\theta}\bigg(\frac{\partial_{\theta}F_2}{\partial_{\theta}F_1}\bigg)\bigg\rbrack},
\eea
where
\bea
\tilde{\lambda}_1\equiv \lambda_1^3\big(\partial_{\theta}F_1\big)^2; \qquad \tilde{\lambda}_2\equiv \frac{\lambda_2^3\partial_{\theta}F_3}{\partial_{\theta}F_1}.
\eea
\\

\noindent
Hence the measure of the SL(3) theory is 
\bea
\int dF_1\wedge dF_2\ 
\frac{1}{\big(\partial_{\theta}F_1\big)^2\bigg\lbrack\partial_{\theta}\bigg(\frac{\partial_{\theta}F_2}{\partial_{\theta}F_1}\bigg)\bigg\rbrack}.
\eea

\end{document}